\documentclass[usenatbib]{mn2e}
\usepackage{graphicx}

\def\apj{\rm ApJ}

\def\mnras{\rm MNRAS}

\def\aap{\rm AAP}

\def\prd{\rm PRD}
\def\apss{\rm Ap\&SS}
\def\physrep{\rm Physics Reports}

\def\gax{\mathrel{\raise.3ex\hbox{$>$}\mkern-14mu\lower0.6ex\hbox{$\sim$}}}
\def\lax{\mathrel{\raise.3ex\hbox{$<$}\mkern-14mu\lower0.6ex\hbox{$\sim$}}}
\def\gtorder{\mathrel{\raise.3ex\hbox{$>$}\mkern-14mu
             \lower0.6ex\hbox{$\sim$}}}
\def\ltorder{\mathrel{\raise.3ex\hbox{$<$}\mkern-14mu
             \lower0.6ex\hbox{$\sim$}}}

\begin{document}

\title [Constraints on Core Collapse]
   {Constraints on Core Collapse from the Black Hole Mass Function}

\author[C.~S. Kochanek]{ C.~S. Kochanek$^{1,2}$\\
  $^{1}$ Department of Astronomy, The Ohio State University, 140 West 18th Avenue, Columbus OH 43210 \\
  $^{2}$ Center for Cosmology and AstroParticle Physics, The Ohio State University,
    191 W. Woodruff Avenue, Columbus OH 43210 
   }

\maketitle

\begin{abstract}
We model the observed black hole mass function under the assumption 
that black hole formation is controlled by the compactness of the 
stellar core at the time of collapse. Low compactness stars are more likely
to explode as supernovae and produce neutron stars, while high compactness stars
are more likely to be failed supernovae that produce black holes with 
the mass of the helium core of the star.  Using three 
sequences of stellar models and marginalizing over a model for the 
completeness of the black hole mass function, we find that the
compactness $\xi_{2.5}$ above which 50\% of core collapses produce 
black holes is $\xi_{2.5}^{50\%}=0.24$ ($0.15 < \xi_{2.5}^{50\%} < 0.37$ at 90\%
confidence).  While models with a sharp transition between successful and failed
explosions are always the most likely ($\xi_{2.5}^{min}=\xi_{2.5}^{max}$), 
the width $\xi_{2.5}^{max}-\xi_{2.5}^{min}$ of the transition between
the minimum compactness for black hole formation $\xi_{2.5}^{min}$ and
the compactness $\xi_{2.5}^{max}$ above which all core collapses produce black holes
is not well constrained.  The models also predict that 
$f=0.18$ ($0.09 < f < 0.39$) of core collapses fail assuming
a minimum mass for core collapse of $8M_\odot$.  We tested four
other criteria for black hole formation based on $\xi_{2.0}$ and $\xi_{3.0}$,
the compactnesses at enclosed masses of $2.0$ or
$3.0$ rather than $2.5M_\odot$, the mass of the iron core,
and the mass inside the oxygen burning shell.  We found that
$\xi_{2.0}$ works as well as $\xi_{2.5}$, while the compactness 
$\xi_{3.0}$ works significantly worse, as does using the
iron core mass or the mass enclosed by the oxygen burning shell.
As expected from the high compactness of $20$-$25M_\odot$ stars,
black hole formation in this mass range provides a natural 
explanation of the red supergiant problem. 
\end{abstract}


\section{Introduction}
\label{sec:introduction}

The mechanism of core collapse supernovae (ccSNe) remains an open problem
despite over half a century of theoretical effort.  In particular, theory
cannot reliably predict which stars that undergo core collapse become
ccSNe or which form black holes instead of neutron stars (e.g.
\citealt{Zhang2008}, \citealt{Nordhaus2010}, \citealt{Oconnor2011}, \citealt{Fryer2012}, 
\citealt{Hanke2012}, \citealt{Takiwaki2012}, \citealt{Ugliano2012}, \citealt{Couch2013}, 
\citealt{Dolence2013}, \citealt{Couch2014}, \citealt{Dolence2014}, \citealt{Wong2014}).
Observationally, comparisons
of massive star formation and SNe rates (\citealt{Horiuchi2011},
\citealt{Botticella2012}), limits on the diffuse SN neutrino background
(\citealt{Lien2010}, \citealt{Lunardini2009}), the Galactic rate (\citealt{Adams2013}),
 and direct searches
for failed SNe (\citealt{Kochanek2008}, \citealt{Gerke2014}) limit
the fraction of failed ccSNe to $f \ltorder 50\%$ of core collapses.
There does, however, appear to be a deficit of higher
mass ccSN progenitors (\citealt{Kochanek2008}), which is best 
quantified by the absence of red supergiant progenitors above
$M \gtorder 17M_\odot$ (\citealt{Smartt2009}).  A natural 
explanation of this deficit is that $f \sim 20$ to $30\%$ of
core collapses fail to produce a ccSNe and instead form a 
black hole without a dramatic external explosion.  Further
progress in completing the mapping between progenitors and
outcomes both for successful ccSNe and in searches for failed
ccSNe should clarify these issues over the next decade.  

We have few other probes of the ccSN mechanism other than this
mapping between progenitors and external outcomes, although
even this does not constrain the balance between neutron star
and black hole formation in successful ccSNe.  Neutrinos or 
gravity waves would be the best probe of the physics of core 
collapse, but this will only be possible for a Galactic supernova
where the rates are low (once every $\sim 50$~years, \citealt{Adams2013}).
Furthermore, the stellar mass function favors having a relatively
low mass progenitor for which the neutrino mechanism for ccSNe
works reasonably well and we have high confidence that the outcome
is a neutron star (e.g., \citealt{Thompson2003}, \citealt{Kitaura2006}, \citealt{Janka2008}) 
rather than a rarer, higher mass progenitor where
the explosion mechanism and the type of remnant remains problematic.  
Even if the rate of failed ccSNe is $f\simeq 20$ to $30\%$, the probability of
detecting the formation of a black hole in the Galaxy is very low.  

Another direct probe of the SN mechanism is the mass function of the remnant
neutron stars and black holes (see, e.g., \citealt{Bailyn1998}, 
\citealt{Ozel2010}, \citealt{Farr2011}, \citealt{Kreidberg2012}, 
\citealt{Ozel2012}, \citealt{Kiziltan2013}).  Most neutron star masses are clustered
around $1.4 M_\odot$,  black hole masses are clustered around
$5$ to $10M_\odot$, and there is a gap (or at least a deep
minimum) between the masses of neutron stars and black 
holes.   Interpreting these results is challenging because
these masses can only be measured in binaries and the selection
functions for finding neutron star and black hole binaries are both
different and difficult or impossible to model from first
principles.  

If, however, we assume that the separate mass distributions of neutron
stars and black holes are relatively unbiased, then we can use them
to constrain the physics of core collapse.  For example, in \cite{Pejcha2012},
we showed that the masses of double neutron star binaries strongly
favored explosion models with no mass falling back onto the proto-neutron
star and that the explosion probably develops at the edge of the
iron core near a specific entropy of $S/N_A \simeq 2.8 k_B$.  Fall-back
is traditionally invoked in order to explain how the observed masses
of black holes can be less than the typical masses of their progenitors
(e.g., \citealt{Zhang2008}, \citealt{Fryer2012}, \citealt{Wong2014}).
However, in \cite{Kochanek2014}, we pointed out that failed ccSNe
of red supergiants naturally produce black holes with the observed
masses because the hydrogen envelope is ejected to leave a remnant
that has the mass of the helium core (\citealt{Nadezhin1980}, 
\citealt{Lovegrove2013}).  Similarly, \cite{Burrows1987} noted
that stellar mass loss processes leads to stars (e.g., Wolf-Rayet stars)
that will produce black holes with masses comparable to that of
the helium core.

Based on these concepts, \cite{Clausen2014} associated the black hole mass with the helium
core mass and then estimated the probability of black hole formation
as a function of progenitor mass needed to explain the observed 
black hole mass function.  As expected from the arguments in
\cite{Kochanek2014}, this required a peak in the probability
distribution at initial (ZAMS) masses of $M_0=20$-$25M_\odot$ 
while also allowing a second peak at $M_0\sim 60M_\odot$ because 
mass loss leads the \cite{Woosley2007} progenitor models to produce
the same helium core mass for two different initial masses. 
However, while stellar mass largely determines the fate of a star, it
is not directly related to the physics of core collapse.  
 
A more physical approach would be to relate the formation of a black
hole to some property of the stellar core at the onset of collapse
that is related to the likelihood of a successful explosion.  This
should not only lead to a more realistic model of the remnant mass
distribution, but the parameter range needed to explain the mass
function can then be used to inform models of core collapse.
\cite{Oconnor2011} argued that the compactness of the core defined by 
\begin{equation}
       \xi_M = { M \over M_\odot} {1000~\hbox{km} \over R(M_{bary}=M) } 
    \label{eqn:compact}
\end{equation}
is a good metric for the ``explodability'' of a star. In particular,
$\xi_{2.5}$, the compactness at a baryonic mass of  $M_{bary}=2.5M_\odot$, is a measure
of how rapidly the density is dropping outside the iron core.  If
$\xi_{2.5}$ is small, the density is dropping rapidly and it is easier
for neutrino heating or other physical effects to revive the shock and
produce a successful explosion.  The reverse holds if the
compactness is high.  \cite{Ugliano2012} argued for a lower
compactness threshold than \cite{Oconnor2011}, and also
considered the correlation of other properties of the core 
with the production of a successful explosion, finding
relatively strong correlations with the binding energy of the
material outside the iron core and little correlation with the
mass of the iron core or the mass inside the oxygen burning shell. 
The compactness is not a simple function of progenitor mass,
and \cite{Sukhbold2014} find that complex interactions between
(in particular) carbon and oxygen burning shells can drive rapid
variations in compactness with stellar mass.

In this paper, following the approach of \cite{Pejcha2012} for
neutron stars, we model the observed black hole mass function
to determine what explosion criteria will explain the black hole
mass function  under the assumption that the black hole mass equals the 
helium core mass at the time of explosion.  While partly inspired
by \cite{Clausen2014}, our approach ties the model to the
physics underlying the success of an explosion rather than
the initial stellar mass.  Unlike 
\cite{Clausen2014}, we will also directly fit the data on
black hole masses from \cite{Ozel2010} rather than fitting
their parametrized model of the black hole mass function.  This is of some importance because
the compactness is a complex function of initial mass, leading
to a mass function with non-trivial structure.  In \S2 we
summarize our statistical model, in \S3 we present our
results, and in \S4 we discuss the future of this approach.

\section{Statistical Methods}
\label{sec:constraints}

There are three elements to our calculation.  First, we must estimate the probability
$P(D_i|M_{BH})$ of the data $D_i$ for black hole candidate system $i$ given an estimated 
black hole mass $M_{BH}$.  Second, we must estimate the probability $P_j(M_{BH}| \vec{p})$
of finding a black hole of mass $M_{BH}$ given a set of model parameters $\vec{p}$ 
describing the outcomes of core collapse for a set of models $j$.  Third, we must 
have some set of priors $P(\vec{p})$ on the model parameters.  
Combining these terms using Bayes theorem, the probability distribution for our model 
parameters given the data on black holes is
\begin{equation}
    P_j(\vec{p}|D) \propto P(\vec{p}) \Pi_i \int dM_{BH} P(D_i|M_{BH}) P_j(M_{BH}|\vec{p}).
\end{equation}
If we only want to consider the probability distribution for the parameters of a
particular model $j$, we simply normalize this distribution to unity.
We can also compare different stellar models or criteria for black
hole formation, where the probability of model $j$
compared to all other models is
\begin{equation}
    P_j(D) = \int d\vec{p} P_j(\vec{p}|D) \left[ \sum_j \int d\vec{p} P_j(\vec{p}|D)\right]^{-1} 
    \label{eqn:relprob}
\end{equation}
for models with the same numbers of parameters.
This is basically the procedure that has been used extensively
to estimate the intrinsic black hole mass distribution (\citealt{Bailyn1998}, \citealt{Ozel2010}, 
\citealt{Farr2011}, \citealt{Kreidberg2012}, \citealt{Ozel2012}, \citealt{Kiziltan2013}) but modified 
as done in \cite{Pejcha2012}  to relate the data to an underlying model of core 
collapse rather than to a parametrized model of the remnant mass distribution. 

For modeling the data, we simply follow the procedures and data summaries from 
\cite{Ozel2010} with one exception.  The probability $P(D_i|M_{BH})$ depends on the data available
for each system.  If both the mass ratio and inclination of a system are constrained,
the probability distribution is described as a Gaussian
\begin{equation}
       P(D_i |M_{BH}) \propto \exp\left[-(M_{BH}-M_i)^2/2\sigma_i^2\right]
     \label{data:case1}
\end{equation}
where $M_i$ and $\sigma_i$ are the estimated mass and its uncertainty and 
the term is always normalized such that $\int P(D_i |M_{BH}) dM_{BH}\equiv 1$.
If the measured mass function is $m_i$ with uncertainty $\sigma_{mi}$
and the mass ratio $q$ is restricted to the
range $q_{min} < q < q_{max}$ then
\begin{equation}
       P(D_i |M_{BH}) \propto \int_{q_{min}}^{q_{max}} dq \int_{x_m}^1 
   { dx \exp\left[-(m_i-m))^2/2\sigma_{mi}^2\right] \over 1-x_m} 
     \label{data:case2}
\end{equation}
where $m=M_{BH}\sin^3 i/(1+q)^2$, the inclination distribution is assumed 
to be uniform in $x=\cos i$ over $x_m < x < 1$, and the  minimum inclination 
$x_m = 0.462 (q/(1+q))^{1/3}$ is set by the requirement for having no
eclipses.  Finally, if there is also a constraint on the inclination,
we include a multiplicative probability for the inclination, 
$\exp(-(i-i_0)^2/2\sigma_i^2)$.  We could reproduce the results in
\cite{Ozel2010} if we also truncated the distributions at $M_{BH}=50M_\odot$.

We model the probability of observing a black hole of mass $M_{BH}$ as
\begin{equation}
    P(M_{BH}|\vec{p}) \propto
      \sum_i { dN \over dM_{0} } \left| dM_{0} \over dM_{BH} \right|_i P(\xi(M_{0})) C(M_{BH})
      \label{eqn:mapping}
\end{equation}
which is also normalized to unity, $\int P(M_{BH}|\vec{p}) dM_{BH} \equiv 1$.
The first term is a Salpeter progenitor mass function,  $dN/dM_{0} \propto M_{0}^{-2.35}$.
The stellar models define a mapping $M_{BH}(M_0)$ between the initial mass and the 
helium core mass we use for the mass of the black hole.  The second term comes
from the variable transformation from $M_0$ in the progenitor mass function to
$M_{BH}$ in the black hole mass function.  The effects of mass loss on higher
mass stars means that the same black hole mass can result for two
different progenitor masses, and we must sum over all solutions $i$.  

The third
term is the probability that a progenitor with some physical property $\xi(M_0)$
will form a black hole.  This is a one-dimensional sequence of a variable
like the compactness and we assume a simple model where $P(\xi)=0$ for
$\xi<\xi^{min}$, $P(\xi)=1$ for $\xi>\xi^{max}$ and that $P(\xi)$ increases
linearly over $\xi^{min} < \xi < \xi^{max}$.  It is also useful to 
constrain $\xi^{50\%}=(\xi^{min}+\xi^{max})/2$, the value at which 50\%
of core collapses produce black holes.  
We might expect a sharp transition
with $\xi^{min}=\xi^{max}$ at which stars either form black holes or
not.  However, there are many secondary variables (e.g., rotation, 
composition, binary mass transfer) that affect stellar evolution beyond mass, 
and stars of a given initial mass likely end with a distribution of 
compactnesses (or any other collapse criterion) at death. It is reasonable
to assume this distribution is largely a spread in final compactnesses
around the values for a particular sequence of progenitor models with mass,
with the net effect of producing a smoothed $P(\xi)$ 
even if the true transition is sharp.  \cite{Clausen2014} proposed using
a mass-dependent probability of black hole formation for similar physical
reasons.

The final term, $C(M_{BH})=(M_{BH}/10 M_\odot)^\alpha$, models the completeness
of the observed black hole mass function.  Because $P(M_{BH}|\vec{p})$ is normalized 
to unit total probability (or, equivalently, that we have no constraint on
the absolute number of black holes), only the shape and not the normalization
of $C(M_{BH})$ affects the results.  If $\alpha >0$ we are more likely
to find high mass black holes, and the reverse if $\alpha <0$.  Completeness
and biases have always been a significant concern for interpreting the
black hole mass function, particularly because black hole masses are only
measured in interacting binaries (see, \citealt{Bailyn1998}, \citealt{Ozel2010}, 
\citealt{Farr2011}, \citealt{Kreidberg2012}, \citealt{Ozel2012}, 
\citealt{Clausen2014}).  Whatever biases exist, they are probably a
relatively smooth function of mass and including $C(M_{BH})$ allows
us to test for their effects or to simply marginalize over
$\alpha$ as a nuisance parameter.

Our model parameters, such as the compactness limits or the exponent
of the completeness function, all have limited dynamic ranges making uniform
priors a reasonable choice.  For comparisons between models with different
parameters (Equation~\ref{eqn:relprob}), we normalize the priors as $\int P(\vec{p}) d\vec{p} \equiv 1$
over the range used for the calculation.  This will give different models
equal relative probabilities in Equation~\ref{eqn:relprob} unless the
data significantly discriminates between the models.  

We also compute
the fraction $f$ of core collapses leading to black holes under the
assumption that all stars with $M_0>8M_\odot$ undergo core collapse. 
We include a weak prior $P(f)$ on the models using the constraints
derived by \cite{Adams2013} from combining the observed Galactic 
SN rate with the lack of any neutrino detection of a 
Galactic black hole formation event.
A stiffer limit of $f < 0.5$ could probably be
justified based on comparisons of SN and star formation rates
(\citealt{Horiuchi2011}) or limits on the diffuse neutrino 
background (\citealt{Lien2010}), but these introduce a dependence
on estimates of star formation rates and are difficult to translate
into a mathematical prior.  These estimates of $f$ are contingent
on the normalization that $P(\xi(M_0))$ becomes unity for 
$\xi > \xi^{max}$.  We could allow $P(\xi(M_0))=P_{max} < 1$
for $\xi > \xi^{max}$ without any consequences for our models
of the black hole mass function, and this would reduce $f$.

While there are differences in the details of the analyses by
\cite{Ozel2010}, \cite{Farr2011} and \cite{Ozel2012}, the 
primary differences in the results are driven by differences
in the samples of black hole candidates.  \cite{Ozel2010} and
\cite{Ozel2012} analyze a sample of 16 systems with low mass
companions, while \cite{Farr2011} analyze 15 low mass systems,
where GC~339$-$4 is the system that is
not in common.  For these low mass systems, the resulting 
inferences about the black hole mass functions are mutually
consistent.  \cite{Farr2011} also carries out the analysis
including 5 systems with high mass companions.  These systems
generally have probability distributions requiring significantly
higher masses than the low mass sample, and 
the inferred black hole mass functions have significantly more
probability for $M_{BH}>10M_\odot$ with their inclusion.  

Like the question of differences in selection functions for
neutron stars and black holes, the relative selection functions for black hole
systems with high and low mass companions probably cannot be derived.
However, the sharply declining mass functions found from
using only the low mass systems are hard to reconcile
with the existence of the high mass systems.
For example, for an exponential mass function, $dN/dM_{BH} \propto \exp(-M_{BH}/M_s)$ with
$ M_c < M_{BH} < 50M_\odot$ fit to the low mass samples,
\cite{Ozel2012} finds $M_c=6.32M_\odot$ and $M_s=1.61M_\odot$ 
while \cite{Farr2011} finds $M_c=6.03M_\odot$
and $M_s=1.55M_\odot$.  These low mass samples exclude
the high mass system Cyg~X1, which has an improved mass 
estimate of $M_{BH}=(14.8\pm1.0)M_\odot$ by \cite{Orosz2011} 
(although \citealt{Ziolkowski2014} argues the uncertainties are somewhat larger).
Given the sample sizes, the probability of finding a system
as massive as Cyg~X1 based on these mass functions is only
about 5\%.  Clearly, adding the information that these 
generally higher black hole mass, high companion mass systems exist will change the inferences
from the low mass systems, just as found by \cite{Farr2011}.

However, of the five high mass systems included by \cite{Farr2011},
only Cyg~X1 is a Galactic source.  If, as seems likely given the
existing samples, the high mass X-ray binaries tend to host higher
mass black holes than the low mass X-ray binaries, then including
the four extragalactic high mass systems without their accompanying
low mass compatriots could be biasing the estimates of the mass
function in the other direction.  As an imperfect compromise,
we model the 16 low mass systems following \cite{Ozel2010} and
include Cyg~X1 with a Gaussian probability distribution of
mean $14.8M_\odot$ and dispersion $2.0M_\odot$ (Equation~\ref{data:case1}), doubling the
uncertainties from \cite{Orosz2011} based on the arguments in
\cite{Ziolkowski2014}.  This gives us a sample of 17 mass 
estimates.

\begin{figure}
\centerline{\includegraphics[width=3.5in]{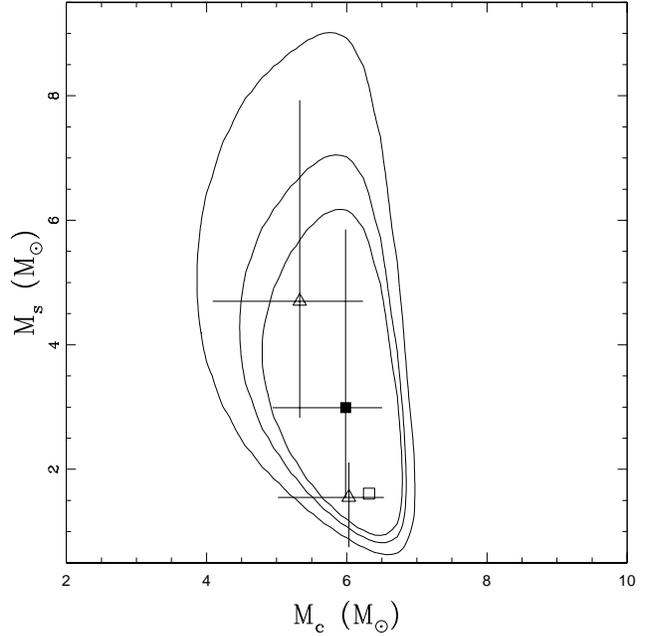}}
\caption{
  Probability contours for the exponential parametric model, $dN/M_{BH}\propto \exp(-M/M_s)$ for
  $M_c < M_{BH} < 50M_\odot$, of the black hole mass function.  The probability contours enclose 90\%, 95\%
  and 99\% of the probability computed over the parameter range shown.  The filled square with
  error bars shows the median value and the 90\% confidence range for each parameter.  The 
  open square with no error bar shows the result from \protect\cite{Ozel2012} using only low mass
  systems.   The open triangles with error bars show the results from \protect\cite{Farr2011}
  for only low mass systems (lower) or both low and high mass systems (upper).  Our
  addition of one high mass system, Cyg~X1, to the \protect\cite{Ozel2010} sample produces
  an intermediate result.
  }
\label{fig:exp}
\end{figure}

In order to compare to the earlier studies, we fit the data using
both an exponential and a power-law ($dN/dM_{BH} \propto \exp(-M_{BH}/M_s)$ or $\propto M_{BH}^\beta$
for $M_c < M_{BH} <50 M_\odot$) parametric mass function.  For
these calculations, Equation~\ref{eqn:mapping} is simply replaced
by the appropriate parametric form.  We used uniform priors for
the two parameters in each models.  Figure~\ref{fig:exp} shows the
results for the exponential mass function, where we find a median
cutoff mass of $M_c=5.98M_\odot$ ($4.94 M_\odot < M_c < 6.51 M_\odot$)
and an exponential scale mass of $M_s=2.99M_\odot$ 
($1.55 M_\odot < M_s < 5.85M_\odot$) where we always present 90\%
confidence intervals.  As expected from adding Cyg~X1, we find a
exponential scale mass that is larger than the results from 
\cite{Ozel2010}, \cite{Farr2011} and \cite{Ozel2012} using only
the low companion mass sample, but smaller than the results from \cite{Farr2011}
including all the high companion mass systems.  For the power law model
we find $M_c=6.21M_\odot$ ($5.57M_\odot < M_c < 6.62 M_\odot$)
and $\beta=-4.87$ ($-7.93 < \beta < -3.00$).  Unlike \cite{Farr2011},
we used a fixed upper mass cutoff at $M_{BH}=50M_\odot$, but the \cite{Farr2011} estimates
of $M_c=6.10 M_\odot$ ($1.28 M_\odot < M_c < 6.63M_\odot$ with 
$\beta=-6.39$ ($-12.42 < \beta < 5.69$) for their low mass
sample and $M_c=5.85 M_\odot$ ($4.87 M_\odot < M_c < 6.46M_\odot$)
with $\beta=-3.23$ ($-5.05 < \beta < -1.77$) for their
combined sample appear compatible with our estimates. 
The relative probabilities
of these two models are about $1.4$ in favor of the exponential
model, but this is well within the regime where the results will
be dominated by the effects of priors (through the choice of the
parameter range over which we carried out the probability integrals).    
 
\begin{figure}
\centerline{\includegraphics[width=3.5in]{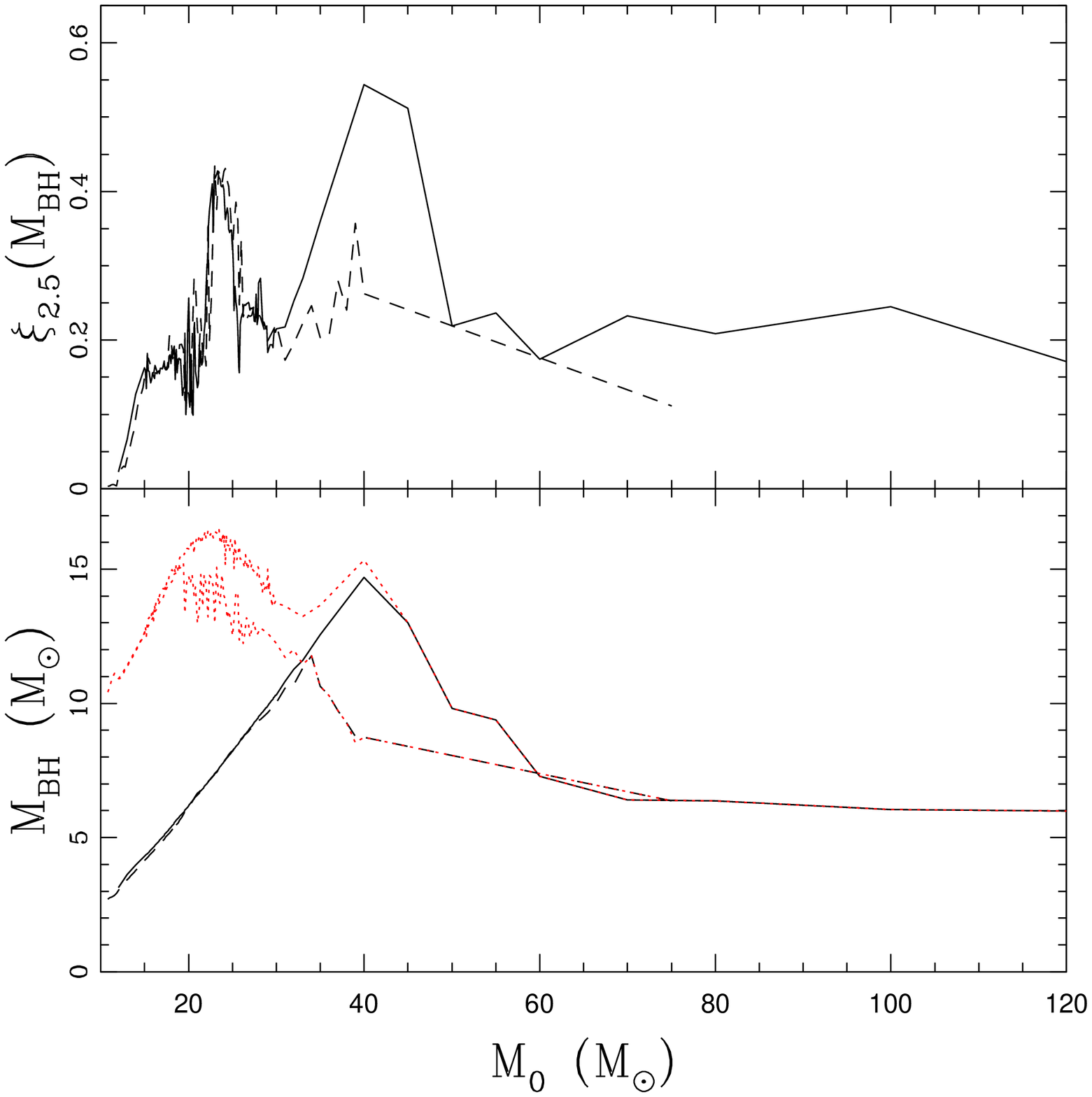}}
\caption{
  Compactness $\xi_{2.5}$ (top) and black hole mass (bottom) as a function of the initial stellar
  mass.  The solid lines are for the \protect\cite{Woosley2007} and \protect\cite{Sukhbold2014} models while the
  dashed lines are for the \protect\cite{Woosley2002} models.  In the lower panel, dotted red lines show
  the final masses of the star.  The hydrogen mass at death (the mass difference between the 
  total mass and the helium core/black hole mass) is assumed to be ejected by the \protect\cite{Nadezhin1980}
  mechanism in a failed supernova.  
  }
\label{fig:profile0}
\end{figure}

\begin{figure}
\centerline{\includegraphics[width=3.5in]{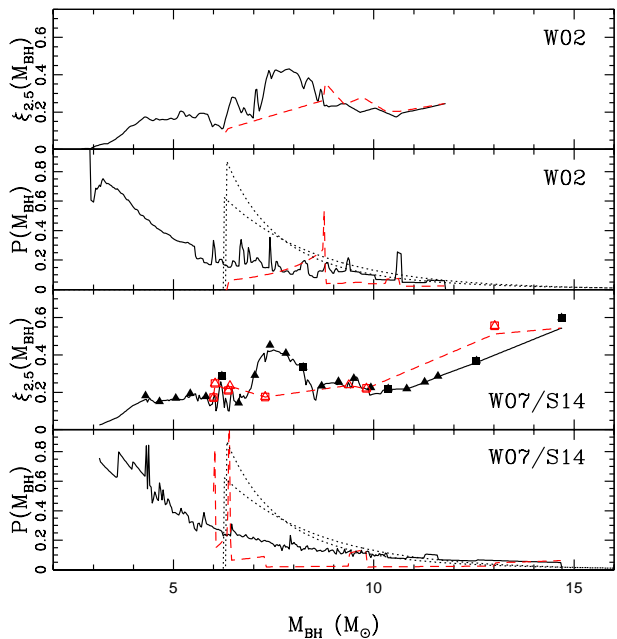}}
\caption{
  Compactness $\xi_{2.5}$ as a function of black hole mass and the black hole mass function
  if all stars formed black holes with the masses of their helium cores for the W02
  (top two panels) and W07$+$S14 (bottom two panels) models.  The solid black (dashed
  red) lines show the contributions from progenitors below (above) the stellar mass producing
  the peak helium core mass.  The normalizations of the mass functions are arbitrary.  
  The squares and triangles in the 
  compactness panel for the W07$+$S14 models show
   the compactness estimates for the same cases from \protect\cite{Oconnor2011}
  and \protect\cite{Sukhbold2014}, respectively.
  Filled black (open red) symbols correspond to low (high) mass progenitors. 
  As noted by \protect\cite{Ugliano2012} there is little difference between the compactness estimates.
  The mass function panels include the parametric mass functions with the median
  parameter estimates from \S2 as the dotted curves.
  }
\label{fig:profile}
\end{figure}

\begin{figure}
\centerline{\includegraphics[width=3.5in]{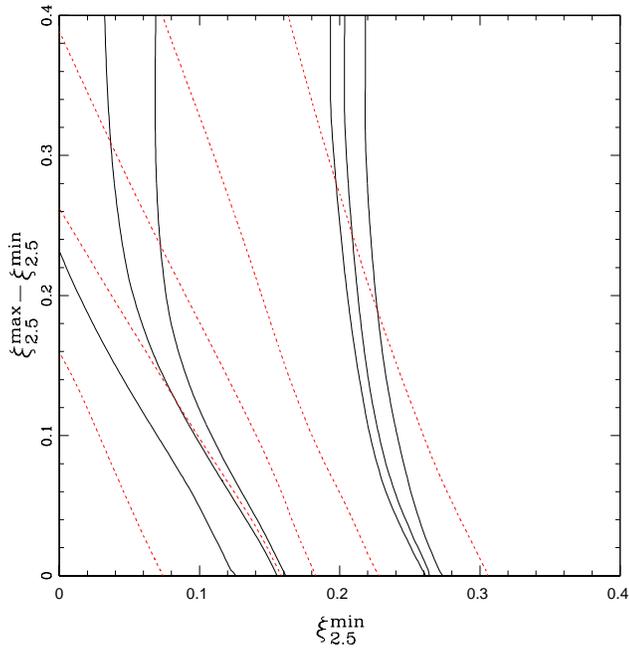}}
\caption{
  Constraints on the compactness $\xi_{2.5}$ for the W07 model assuming no biases
  in the black hole mass function ($\alpha=0$).  The probability contours
  enclose 90, 95 and 99\% of the total probability for a uniform prior 
  over the region shown.  The red dotted lines show
  contours of the failed ccSNe fraction with $f=0.1$, $0.2$, $0.3$, $0.4$ and $0.5$ 
  (from right to left). 
  }
\label{fig:xi1}
\end{figure}

\begin{figure}
\centerline{\includegraphics[width=3.5in]{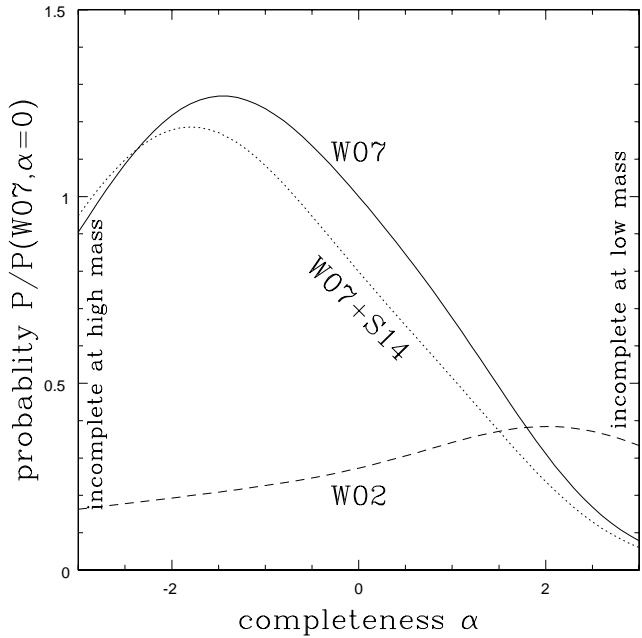}}
\caption{
  Model probabilities as a function of the completeness exponent $\alpha$ for the
  W02 (dashed), W07 (solid) and W07$+$S14 (dotted) $\xi_{2.5}$ models.  The 
  probabilities are relative to the $\alpha=0$ W07 model.  The mass function
  is incomplete at high mass for $\alpha<0$ and incomplete at low mass for
  $\alpha > 0$.
  }
\label{fig:incomp}
\end{figure}

\begin{figure}
\centerline{\includegraphics[width=3.5in]{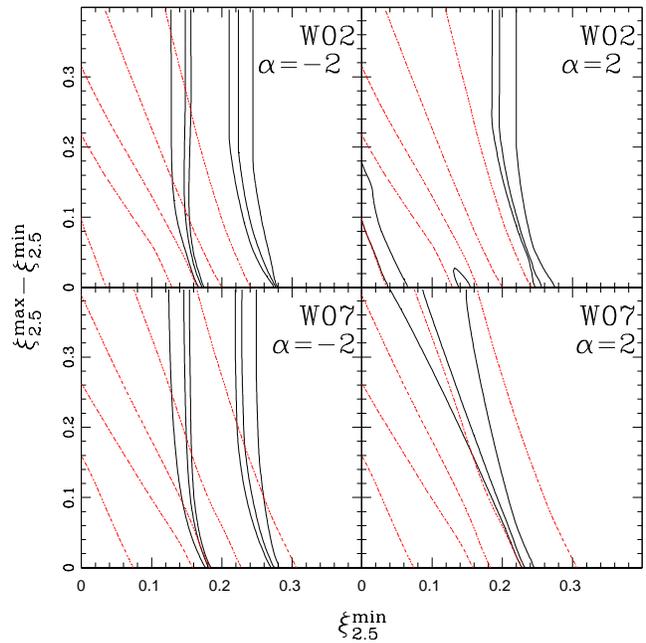}}
\caption{
  Constraints on the compactness $\xi_{2.5}$ as a function of the completeness.
  The top panels are for the W02 model and the bottom panels are for the W07
  model.  The left panels assume the observations are more complete for low
  mass black holes ($\alpha=-2$) while the right panels assume the observations
  are more complete for high mass black holes ($\alpha=2$).   Low values of
  $\xi_{2.5}^{min}$ are favored in the $\alpha=2$ models.  The probability contours 
  enclose 90, 95 and 99\% of the total probability for each model.
  The red dotted lines show
  contours of the failed ccSNe fraction with $f=0.1$, $0.2$, $0.3$, $0.4$ and $0.5$ 
  (from right to left). 
  }
\label{fig:xi2}
\end{figure}

\begin{figure}
\centerline{\includegraphics[width=3.5in]{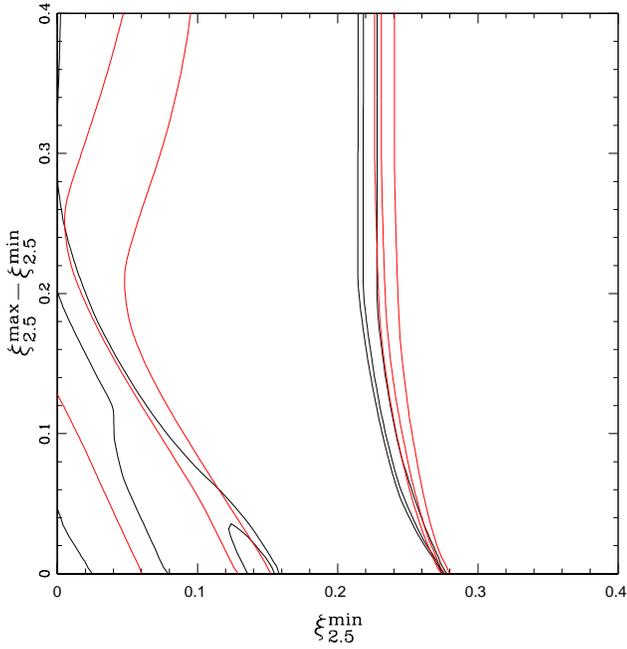}}
\caption{
  Constraints on the compactness $\xi_{2.5}$ marginalized over completeness ($-3 < \alpha < 3$)
  for the W02 (black solid) and W07 (red dotted) models.  The results for the W07$+$S14
  models are very close to those for the W07 models.  The probability contours enclose
  90, 95 and 99\% of the total probability for each model.  
  }
\label{fig:xi3}
\end{figure}

\begin{figure}
\centerline{\includegraphics[width=3.5in]{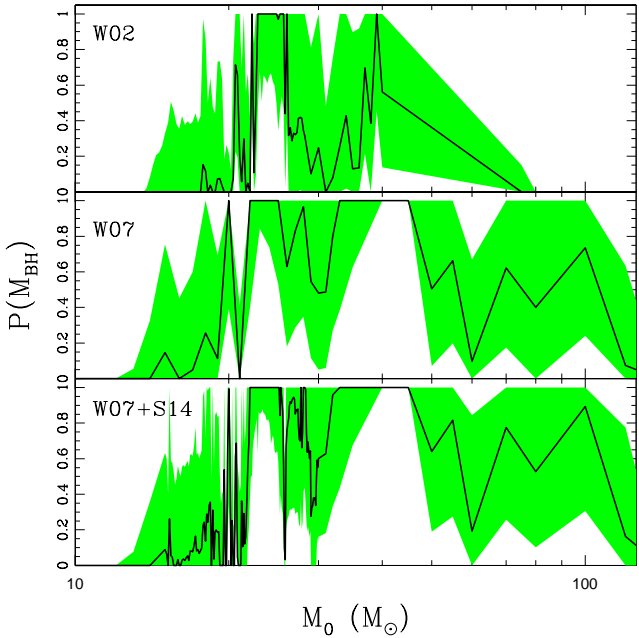}}
\caption{
  Probability of black hole formation as a function of initial stellar mass for $\alpha=0$
  and the W02 (top), W07 (middle) and W07$+$S14 (bottom) models after marginalizing 
  over $\xi_{2.5}^{min}$ and $\xi_{2.5}^{max}$.  The heavy black line is the median probability
  and the shaded band is the 90\% confidence range.  The estimates are highly correlated
  between different masses. 
  }
\label{fig:prob}
\end{figure}

\begin{figure}
\centerline{\includegraphics[width=3.5in]{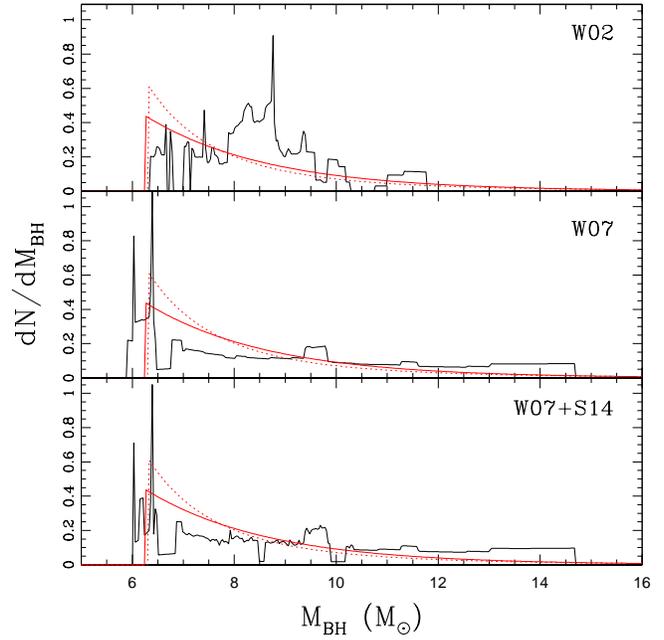}}
\caption{
  Maximum likelihood mass functions for the W02 (top), W07 (middle) and W07$+$S14 
  models with $\alpha=0$.  They are normalized to have the same integrated area.
  For comparison, the red solid and dashed curves show the parametric 
  mass function estimates from \S2.  Most of the structures here are real
  and are created by the rapid changes of the compactness with $M_0$ in
  some mass ranges (see \protect\citealt{Sukhbold2014}). 
  }
\label{fig:mfunc}
\end{figure}

\section{Results}
\label{sec:results}

We consider the \cite{Woosley2002} and \cite{Woosley2007} Solar metallicity progenitor models.
The \cite{Woosley2002} models span $10.8 M_\odot < M_0 < 70 M_\odot$ with a relatively
dense sampling ($\Delta M=0.2 M_\odot$) from $10.8 M_\odot$ to $30 M_\odot$, a 
coarser sampling ($\Delta M=1.0 M_\odot$) to $40M_\odot$ and then a last model
at $75 M_\odot$.  The \cite{Woosley2007} models span $12 M_\odot < M_0 < 120M_\odot$
sampling every $\Delta M =1.0 M_\odot$ up to $35 M_\odot$, every $\Delta M=5.0M_\odot$
to $60 M_\odot$, and models at $70$, $80$, $100$ and $120 M_\odot$.  \cite{Sukhbold2014}
supplemented the \cite{Woosley2007} models, sampling the range from $15$ to $30M_\odot$ with $\Delta M= 0.1 M_\odot$.
We assume models from $8M_\odot$ to $75 M\odot$ for the \cite{Woosley2002} models or
$120 M_\odot$ for the \cite{Woosley2007} models undergo core collapse and allow black
hole formation only over the mass range of the model sequences.  Stars with masses
from $8M_\odot$ to the minimum masses of the sequences ($10.8$ or $12M_\odot$) are 
assumed to be successful SNe forming neutron stars. We will refer
to the three progenitor sequences as the W02, W07, and W07$+$S14 models.

For each model we computed the mass of the helium core, defined by the radius where
$X_H = 0.2$, and the compactness $\xi_M$ (Equation~\ref{eqn:compact}) 
for $M_{bary}=2.0$, $2.5$ and $3.0$ based on the
progenitor model.  For the densely sampled mass ranges we slightly smoothed the
helium core masses as a function of mass to remove small fluctuations that
complicate the mapping from initial mass into black hole mass due to the derivative
in Equation~\ref{eqn:mapping}.  Following \cite{Ugliano2012} we also computed
the mass of the iron core ($M_{Fe}$, defined by the point where $Y_e=0.497$) and
the mass at the inner edge of the oxygen burning shell ($M_O$, defined by the point
where the dimensionless entropy per baryon equaled 4) for the W02 and W07
models.  We did not possess the density profiles needed to compute these
properties for the S14 models, but \cite{Sukhbold2014} 
kindly supplied a table of their helium core
masses and $\xi_{2.5}$.  

\cite{Oconnor2011} argue that $\xi_{2.5}$
should be computed at the time of core bounce, while \cite{Sukhbold2014} note
that computing it when the collapse speed reaches $1000$~km/s produces
equivalent results and \cite{Ugliano2012} note that there is little
difference between simply calculating it from the progenitor model and
calculating it at the time of core bounce. Operationally, simply using the 
compactness of the progenitor is far simpler because it avoids
simulating a portion of the collapse for each model.
If we use the \cite{Oconnor2011} or \cite{Sukhbold2014} values, we
find no significant differences from our results simply using the values estimated
from the progenitors.  

Figure~\ref{fig:profile0} shows the helium core
mass, which we now define to be the black hole mass, and the compactness
$\xi_{2.5}$ as a function of the initial stellar mass $M_0$.  The main qualitative
difference is that the \cite{Woosley2002} models have a lower maximum helium
core mass and a significantly smaller and lower compactness peak near $M_0 \simeq 40 M_\odot$.
The other compactness sequences look similar but with modest changes in the
average level and the ratio of the peaks near $20$-$25$ and $40 M_\odot$.

Figure~\ref{fig:profile0} also shows the total mass when the star explodes,
and we assume that any remaining hydrogen envelope is ejected.  For a failed
supernova of a red supergiant, which represents the bulk of any models with
residual hydrogen, the \cite{Nadezhin1980} mechanism naturally does so,
as seen in the simulations of \cite{Lovegrove2013}.  For the higher mass stars,
mass loss has stripped the hydrogen prior to the explosion.  As the
amount of residual hydrogen becomes negligible, the stellar envelope will
collapse and the \cite{Nadezhin1980} mechanism cannot work because it depends on the
very low binding energy of a red supergiant envelope. At this point, however,
the correction from adding the remaining hydrogen to the black hole is 
also unimportant.
While we are
phrasing black hole formation as being due to failed SNe, fine tuning fall
back could produce the same black hole masses in successful SNe.  

Figure~\ref{fig:profile} shows the structure of these models as a function
of black hole mass.  Because the black hole mass peaks at an intermediate
initial mass, there are two branches to the solutions, corresponding to
initial masses above and below this maximum.  Broadly speaking, there
is an extended plateau in $\xi_{2.5}$ starting near $M_{BH}\simeq 4M_\odot$
with a rapid drop at lower masses, a narrow peak near $6 M_\odot$ and
a broader peak near $8M_\odot$.  We also show the $\xi_{2.5}$ values
from \cite{Oconnor2011} and \cite{Sukhbold2014}, and we see that there
are few differences from our estimates simply using the structure of
the progenitor, as previously noted by \cite{Ugliano2012}.

We also show the black hole mass functions
which would result from all the models becoming black holes with the mass
of the helium core.  The general decline of the Salpeter mass function
is visible, but the slope of $M_{BH}(M_0)$ introduces significant structure.
The broader trends are real features of the models, but some of the small scale
structure is due to ``noise'' in $M_{BH}(M_0)$.  Whenever $M_{BH}(M_0)$
is locally flat, the derivative term in Equation~\ref{eqn:mapping} leads
to a local peak in the mass function.   The smoothing we used to make the
$M_{BH}(M_0)$ profiles monotonic outside of the single peak significantly reduces these
features compared to the raw relations.  None of these local structures
are important because they are changes in the mass function on such small
scales that they have no consequences for modeling the data.  For comparison,
we also show the two parametric estimates of the black hole mass function from
\S2.  Producing a similar mass function
requires suppressing the formation of low mass black holes.  The mass 
functions derived from the stellar models do not decline as rapidly 
as the parametric models, which could be a problem in either of the models
or a consequence of completeness in the black hole sample.

We first consider the three models using $\xi_{2.5}$ and no mass-dependent
completeness effects ($\alpha=0$).  These W02, W07 and W07$+$S14 models have
probabilities relative to the exponential parametric model of  $0.7$,
$2.5$ and $2.0$, respectively.  Thus, our first result is that our 
models based on progenitor models and a physical criterion for 
black hole formation can fit the observed mass function just as well 
as standard parametric models.  However, given the changing parameters
and parameter ranges, the differences are not large enough to argue 
that the $\alpha=0$ W07 model is significantly better.  Between the
three stellar models, the ratios are $0.27:1.0:0.80$, favoring the
W07 model but not by a large enough factor to rule out the W02 model.
This ordering is driven by the inclusion of Cyg~X1.  If we drop it,
the relative probabilities are $1:0.33:0.26$ and the W02 model is
favored over the W07 model.  These shifts are driven by the relative
areas of the compactness peaks in the two models: with Cyg~X1, W02
produces too few massive black holes, and without Cyg~X1, W07 produces
too many.   We will explore this in more detail when we discuss the
effects of the completeness model. 

Figure~\ref{fig:xi1} shows the likelihood function for the $\alpha=0$,
$\xi_{2.5}$ W07 model as
well as the fractions of core collapses becoming black holes. The structures
of the W02 and W07$+$S14 results are similar.  The compactness threshold
$\xi_{2.5}^{min}$ for black hole formation is well constrained. Although the
most probable model is always for a sharp transition with $\xi_{2.5}^{max}=\xi_{2.5}^{min}$,
the width of the transition, $\xi_{2.5}^{max}-\xi_{2.5}^{min}$, is not well constrained.
Formally, the median estimates are $\xi_{2.5}^{min}=0.17$ ($0.07 < \xi_{2.5}^{min} < 0.23$),
$\xi_{2.5}^{min}=0.16$ ($0.06 < \xi_{2.5}^{min}<0.22$), and
$\xi_{2.5}^{min}=0.15$ ($0.06 < \xi_{2.5}^{min}<0.20$)
for the W02, W07 and W07$+$S14 models, respectively.  This is somewhat
misleading because the lower values of $\xi_{2.5}^{min}$ are associated with
wide transitions.  A better metric is probably $\xi_{2.5}^{50\%}=(\xi_{2.5}^{min}+\xi_{2.5}^{max})$,   
the point where the probability of forming a black hole becomes 50\%.
For these models we find that
$\xi_{2.5}^{50\%}=0.24$ ($0.17<\xi_{2.5}^{50\%}<0.36$),
$\xi_{2.5}^{50\%}=0.23$ ($0.17<\xi_{2.5}^{50\%}<0.33$), and
$\xi_{2.5}^{50\%}=0.21$ ($0.16<\xi_{2.5}^{50\%}<0.32$), so $\xi_{2.5}^{50\%}$  is constrained
to be close the plateau in $\xi_{2.5}$ seen in Figure~\ref{fig:profile}
and high enough to largely prevent black hole formation at lower 
masses where the compactness is dropping rapidly. As we will see,
this is sufficient to lead to a sharp break in the black hole
mass function just as is included in the parametric models.
These estimates from fitting the observed black hole mass
function are quite similar to the range of $0.15$ to $0.35$ 
found by \cite{Ugliano2012} in their core collapse simulations
and lower than the limit proposed by \cite{Oconnor2011}.
The fractions of core collapses producing black holes are
$f=0.13$ ($0.05<f<0.28$),
$f=0.21$ ($0.11<f<0.34$), and 
$f=0.21$ ($0.11<f<0.33$), respectively, where the W07 models
produce larger numbers of black holes because of the prominent compactness
peak near $M_0=40 M_\odot$.  If we arbitrarily remove this
peak by setting $\xi_{2.5}=0.2$ in this regime, the results
more closely resemble those for the W02 models.

For the W02 and W07 models we can also examine using the compactness
at another mass cut, $\xi_{2.0}$ or $\xi_{3.0}$, the iron core mass,
$M_{Fe}$, or the  mass at inner edge of the oxygen burning shell, $M_O$,
as the criterion for forming a black hole.  The procedure is the
same in each case, we raise the probability for forming a black hole
linearly from zero at one value of the parameter to unity at a higher
value.  For the $\alpha=0$ W02 models, the relative probabilities of
the criteria are $1.97:1.00:0.31:0.19:0.20$ for $\xi_{2.0}$, $\xi_{2.5}$,
$\xi_{3.0}$, $M_{Fe}$ and $M_O$, respectively, where we have normalized
the models to the standard $\xi_{2.5}$ model.  For the $\alpha=0$ W07 models,
the ratios are $1.11:1.00:0.52:0.28:0.03$.  This general structure
holds when we vary the completeness as well.  If we simply marginalize
over the completeness ($-3 < \alpha < 3$ with a uniform prior) and
the W02 and W07 models, the probabilities of the formation criteria relative
to the $\xi_{2.5}$ model are $0.92:1.00:0.39:0.29:0.14$. The
compactness at a smaller mass cut $\xi_{2.0}$ is almost equally
good, the compactness at a larger mass cut $\xi_{3.0}$ is moderately
worse, the iron core mass is an even poorer model, and the 
oxygen burning shell mass is the worst model.  Arguably, only
$M_O$ shows a large enough probability ratio to be rejected
at a reasonable confidence level.  These results are consistent
with the simulations of \cite{Ugliano2012}, who found that 
$M_{Fe}$ and $M_O$ had poorer correlations with black hole
formation than $\xi_{2.5}$.  For the remainder of the paper
we will just consider the $\xi_{2.5}$ model.

Figure~\ref{fig:incomp} shows the relative probabilities of the 
W02, W07 and W07$+$S14 $\xi_{2.5}$ models as a function of
the completeness exponent $\alpha$.  The W02 models modestly 
prefer incompleteness at low mass, while the W07 and
W07$+$S14 models more strongly prefer incompleteness at 
high mass.  We only explored the range $-3 < \alpha < 3$
with a uniform prior. All three cases peak in this range
and larger values of $|\alpha|$ seemed unreasonable -- 
for $|\alpha|=3$, the relative completeness changes by a
factor of 8 between $M_{BH}=5$ and $10M_\odot$.  The
qualitative differences can be understood from the 
differences in the $\xi_{2.5}$ profiles shown in 
Figures~\ref{fig:profile0} and \ref{fig:profile}.  
The peak near $M_0=40 M_\odot$ which produces the most
massive black holes is weaker compared to the peak
near $M_0=20$-$25M_\odot$ in the W02 models relative
to the W07 or W07$+$S14 models.  Compared to the data,
the W02 model produces too few high mass black holes
and so to compensate the models make the observed sample 
incomplete at low masses.  The W07 models produce 
too many high mass black holes, and compensate by
making the observed sample incomplete at high mass. 
If we exclude Cyg~X1, then all three models favor
observational samples that are incomplete at high 
black hole mass.  

The completeness probability distributions are related to the results of
\cite{Clausen2014} in their estimate of the probability
of black hole formation as a function of initial stellar
mass $M_0$.  Their formation probability distribution has a peak
associated with the first peak in $\xi_{2.5}$ at
$M_0=20$-$25M_\odot$ but no peak associated with the
second peak in $\xi_{2.5}$ near $M_0=40 M_\odot$.   
They derived the probability distributions by fitting
the parametric mass functions of \cite{Ozel2012}, 
which are based on only the low mass binaries and
have low probabilities for higher mass black holes.
The lack of a probability peak near $M_0=40 M_\odot$
is similar to the effects of $\alpha <0$. 

For comparison to the mass-independent completeness case
in Figure~\ref{fig:xi1}, 
Figure~\ref{fig:xi2} shows the results for the W02 and W07
models with $\alpha=-2$ and $2$.  This corresponds to 
changing the completeness between $M_{BH}=5M_\odot$
and $10M_\odot$ by a factor of 4.   When the discovery
of low mass black holes is favored ($\alpha=-2$), the
upper limits on $\xi_{2.5}^{min}$ change little but the lower
limits become much stronger.  When the discovery of high
mass black holes is favored ($\alpha=2$), the upper limit
on $\xi_{2.5}^{min}$ becomes somewhat stronger, while the lower
limit becomes significantly weaker.  As the completeness
for low mass black holes decreases, there is more freedom
to make them.  This can be seen in the fractions of core
collapses producing black holes, which are $f=0.10$ ($0.04<f<0.20$)   
and $0.08$ ($0.14<f <0.25$) for the W02 and W07 models with 
$\alpha=-2$ but $f=0.20$ ($0.06 < f < 0.45$) and 
$f=0.34$ ($0.18<f<0.51$) for the $\alpha=2$ models.  

Finally, we can also marginalize over the completeness model
($-3 < \alpha < 3$ with a uniform prior).  Marginalized
over $\alpha$, the relative probabilities of the W02, W07 
and W07$+$S14 $\xi_{2.5}$ models are $0.33:1.00:0.88$, moderately 
favoring the W07 models.  Figure~\ref{fig:xi3} shows the
constraints on the compactness parameters for the W02 and
W07 models marginalized over completeness.  The hard upper
limit is robust, but the lower limit becomes soft because
of the contribution from the $\alpha > 0$ models.
The estimates of the critical compactness $\xi_{2.5}^{min}$ are
$\xi_{2.5}^{min}=0.17$ ($0.03 < \xi_{2.5}^{min} < 0.23$),
$\xi_{2.5}^{min}=0.18$ ($0.05 < \xi_{2.5}^{min} < 0.23$), and 
$\xi_{2.5}^{min}=0.17$ ($0.05 < \xi_{2.5}^{min} < 0.21$)
for the W02, W07 and W07$+$S14 models, respectively.
The constraints on $\xi_{2.5}^{50\%}$, the compactness where the
probability of forming a black hole is 50\%, are tighter,
with 
$\xi_{2.5}^{50\%}=0.23$ ($0.11 < \xi_{2.5}^{50\%} < 0.35$),
$\xi_{2.5}^{50\%}=0.24$ ($0.15 < \xi_{2.5}^{50\%} < 0.37$), and
$\xi_{2.5}^{50\%}=0.23$ ($0.14 < \xi_{2.5}^{50\%} < 0.35$).
Finally, the constraints on the fraction of core collapses
leading to black holes are
$f=0.14$ ($0.05 < f < 0.41$),
$f=0.18$ ($0.09 < f < 0.39$), and
$f=0.18$ ($0.09 < f < 0.38$).
Despite allowing for very large, black hole mass-dependent 
completeness corrections, the constraints on $\xi_{2.5}^{50\%}$
and $f$ are quite strong.  

\cite{Clausen2014} estimated the probability of forming a black
hole with the mass of the helium core as a function of initial
stellar mass needed to match the \cite{Ozel2012} parametric 
model of the black hole mass function.  For a given model, we
can marginalize over $\xi_{2.5}^{min}$ and $\xi_{2.5}^{max}$ to estimate
this probability as a function of mass as well as its variance
over the model space, although the uncertainty estimates are
highly correlated.  The results for the $\alpha=0$ 
case are shown in Figure~\ref{fig:prob}.
Black hole formation is disfavored below $M_0 \simeq 20M_\odot$
with the probability steadily dropping towards lower masses.
Black hole formation is probably 
required near $20$-$25M_\odot$ and near $40M_\odot$.  
At most other masses, the data do
not strongly constrain the probability of forming a black 
hole.

Finally, in Figure~\ref{fig:mfunc} we show the maximum likelihood
mass functions for the three $\alpha=0$, $\xi_{2.5}$ models as
compared to the median fit parametric exponential and power-law models
from \S2.  Our models roughly match the minimum 
black hole masses of the parametric models.  The W07 and W07$+$S14 mass
functions are relatively flat due to the broad compactness
peak at $M_0 \sim 40M_\odot$, while the W02 mass function
is more strongly peaked at low masses.  The absence of 
higher mass black holes in the W02 models as compared to
the W07/W07$+$S14 models explains why the W07/W07$+$S14 models 
are favored with the inclusion of Cyg~X1.  

\section{Discussion}
\label{sec:discuss}

If black hole formation is controlled by the compactness of
the stellar core at the time of collapse (e.g., 
\citealt{Oconnor2011}, \citealt{Ugliano2012}, \citealt{Sukhbold2014})
and we associate the mass of the resulting black hole
with the mass of the helium core (e.g., \citealt{Burrows1987},
\citealt{Kochanek2014}, \citealt{Clausen2014}) then we 
can constrain the compactness above which black holes
form by fitting the observed black hole mass function
(e.g., \citealt{Bailyn1998},
\citealt{Ozel2010}, \citealt{Farr2011}, \citealt{Kreidberg2012},
\citealt{Ozel2012}).  The helium core mass is a natural
scale for black hole masses due to either mass loss
(e.g., \citealt{Burrows1987}) or the physics of failed
ccSNe (\citealt{Nadezhin1980},
\citealt{Lovegrove2013}, \citealt{Kochanek2014}).  We
also, for the first time, include a model for the 
completeness of the observed sample of black holes and
examine its consequences for the constraints on the 
core collapse parameters.

We use a sample of 17 black hole candidates, the 16 low
mass binary systems used by \cite{Ozel2010} and \cite{Ozel2012}
combined with the one Galactic high mass system, Cyg~X1.
\cite{Farr2011} found that the parameters of their models
of the black hole mass function changed significantly when
the high mass binaries were included in the analysis because
they tend to have higher average masses.  Including 
Cyg~X1 is a compromise between excluding all high mass
systems (\citealt{Ozel2010}, \citealt{Ozel2012}, \citealt{Farr2011}) and
including all high mass systems (\citealt{Farr2011}). We
exclude the extragalactic high mass systems since there
is no equivalent sample of extragalactic low mass systems.
Where necessary, we discuss the impact of including Cyg~X1
on the results.  If we model the data using the parametric 
methods and models of these previous studies, we obtain 
similar results.  

The first interesting result is that our models based on 
combining progenitor models with a physical criterion for
forming a black hole fit the observed black hole mass function
as well as the existing parametric models.  Our best model actually has
a higher likelihood, but the likelihood ratios are not
large enough to be significant.  Unlike the simple parametric
models, the mass functions produced
by our models have a great deal of structure (Figure~\ref{fig:mfunc})
created by the rapid variations in compactness with
mass (Figures~\ref{fig:profile0} and \ref{fig:profile}). 

We tested five different parameters for predicting the
formation of a black hole.  The compactnesses $\xi_{2.0}$,
$\xi_{2.5}$ and $\xi_{3.0}$ of the core of the progenitor 
at baryonic masses of $2.0$, $2.5$ and $3.0M_\odot$
(Equation~\ref{eqn:compact}), the
mass of the iron core $M_{Fe}$ and the mass inside the
oxygen burning shell $M_O$.  We considered three sets
of stellar models, W02 from \cite{Woosley2002}, 
W07 from \cite{Woosley2007}, and W07$+$S14 which
supplements the W07 models with the denser mass 
sampling from \cite{Sukhbold2014}.  
Marginalizing over all our other variables, we
find that the compactnesses $\xi_{2.0}$ and $\xi_{2.5}$
produced the highest likelihoods.  The compactness
$\xi_{3.0}$ models were somewhat less probable, the
iron core mass models still less so, and the oxygen 
shell mass models were the worst.  With overall 
likelihood ratios relative to the $\xi_{2.5}$ model
of $0.92:1.00:0.39:0.29:0.14$ none of the other 
possibilities are strongly ruled out.  \cite{Ugliano2012}
found in their simulations that the compactness was
a better predictor of outcomes than the iron core or
oxygen shell masses. We now focus
on the $\xi_{2.5}$ models.  

The relative probabilities of the W02, W07
and W07$+$S14 $\xi_{2.5}$ models after marginalizing
over the completeness model are $0.33:1.00:0.88$,
favoring the W07 models.  The origin of the differences
are due to the relative strengths of the peaks in the
compactness as a function of mass near $M_0=20$-$25M_\odot$
and $M_0=40M_\odot$, which control the relative production
of higher and lower mass black holes.  The W02 models
have difficulty producing higher mass black holes and 
so are disfavored.  This estimate is affected
by the inclusion of Cyg~X1 --  if Cyg~X1 is excluded, the
relative probabilities favor the W02 models over the 
W07 or W07$+$S14 models by similar factors.
  
The W07 and W07$+$S14 cases favor models in which the observed
black hole mass function is incomplete at high masses, while
the W02 case favors models in which it is incomplete at low masses.
This is again a reflection of the relative importance of 
the two compactness peaks.  If we exclude Cyg~X1, thereby
dropping the (probably) highest mass black hole in the sample,
all three models favor a black hole mass function that is
incomplete at high masses.  These trends are related to the
estimate of the black hole formation probability as a function
of initial stellar mass by \cite{Clausen2014}.  Using only
the low mass companion systems, their models have a low probability
of black hole formation at the compactness peak near 
$M_0=40 M_\odot$ that produces the higher mass black holes.
This is equivalent to our models reducing the completeness
of the observed sample for higher mass black holes.
That our completeness models generally favor incompleteness
for high mass systems rather than low mass systems supports
the existence of a gap or a deep minimum
between the masses of neutrons stars
and black holes.

Even with a model allowing large variations in the 
completeness as a function of mass, we obtain interesting
constraints on the compactness leading to black hole
formation.  We modeled the probability of black hole
formation as linearly rising from zero at a minimum compactness
$\xi_{2.5}^{min}$ to unity at $\xi_{2.5}^{max}$.  
The probability of black hole
formation should increase monotonically with compactness,
but stars of a given mass will really end their lives
with a range of compactnesses because of secondary 
variables other than their initial mass (e.g., composition,
rotation, binary interactions).  If these secondary
variables produce a spread in the compactness at a given
initial mass,
the finite width of our transition will mimic much of
the effect.  That being said, the maximum likelihood
model was always the one with an abrupt transition
($\xi_{2.5}^{min}=\xi_{2.5}^{max}$), although probability of
$\xi_{2.5}^{max}-\xi_{2.5}^{min}>0$ only declines slowly and
the width of the transition is not well constrained.
For models in which the completeness is mass-independent,
we find $\xi_{2.5}^{min}=0.15$ ($0.06 < \xi_{2.5}^{min}<0.20$)
and $\xi_{2.5}^{50\%}=0.23$ ($0.17<\xi_{2.5}^{50\%}<0.33$) where 
$\xi_{2.5}^{50\%}=(\xi_{2.5}^{min}+\xi_{2.5}^{max})/2$
is the compactness at which the black hole formation 
probability is 50\%.  Because the width of the 
transition is poorly constrained, $\xi_{2.5}^{50\%}$ has
smaller uncertainties than $\xi_{2.5}^{min}$.
If we marginalize over the completeness model,
we find $\xi_{2.5}^{min}=0.18$ ($0.05 < \xi_{2.5}^{min} < 0.23$)
and $\xi_{2.5}^{50\%}=0.24$ ($0.15 < \xi_{2.5}^{50\%} < 0.37$). 
The upper limits change little,
but the lower limits become softer because of the 
contribution from models with high incompleteness at low black
hole mass.  These
results are for the W07 models, but the results for the
other two cases are similar.  

The fraction of core collapses predicted to form  black
holes is relatively high.  For the W07 model with no
mass-dependent completeness corrections, $f=0.21$ ($0.11<f<0.34$),
while after marginalizing over the completeness corrections,
$f=0.18$ ($0.09 < f < 0.39$).  The results for the other
model sequences are similar.  This fraction assumes a
Salpeter IMF where all stars from $M_0=8M_\odot$ to the 
maximum mass in the model sequence undergo core collapse.
It also assumes that the probability of black hole formation
for $\xi_{2.5}>\xi_{2.5}^{max}$ is unity, $P_{max}\equiv 1$.  Lowering
$M_0$ or $P_{max}$ would reduce $f$, while raising $M_0$
would increase $f$.  For example, raising the minimum mass
for core collapse from $8M_\odot$ to $9M_\odot$ raises
$f$ by 18\%.  The probability of black hole formation as a
function of initial stellar mass is a complex function
reflecting the structure of $\xi_{2.5}(M_0)$.  Generically,
the probability drops rapidly below $M_0 =20 M_\odot$ 
in order to minimize the production of low mass black holes
and is unity near $M_0=20$-$25M_\odot$ and $M_0=40M_\odot$
where the compactness peaks.  At other mass ranges, the
probabilities vary greatly with $\xi_{2.5}^{min}$ and $\xi_{2.5}^{max}$.  

The most interesting directions for expansion would be to
consider other potential indicators of outcomes, such as
binding energies, other sequences of stellar models, and
other possible definitions of the black hole mass.  In,
particular, many model sequences of ccSN progenitors span
only portions of the mass range of interest, frequently
with large spacings in mass.  In order to carry out the
analysis, this approach requires the full mass range from
$M_0 \sim 8 M_\odot$ to $M_0 \sim 100 M_\odot$.  While
the differences between the densely sampled \cite{Sukhbold2014}
and the sparser \cite{Woosley2007} models in the mass range
$15M_\odot < M_0 < 30 M_\odot$ seem to have little consequence
for our results, we are concerned that the sparse sampling
of the \cite{Woosley2007} models near $M_0 \sim 40 M_\odot$
compared to the more densely sampled \cite{Woosley2002} models 
may drive some differences in the contribution of stars in
this mass range to the black hole mass function.  Our  
approach, including models of sample completeness,
 could also be used to fit sequences of fall-back
models (e.g., \citealt{Zhang2008}, \citealt{Fryer2012}) to the 
black hole mass function.    We have not done so here because
these fall-back models produce continuous mass functions
that are completely incompatible with the existing data. 

We used a fairly minimalist prior on the fraction of failed
ccSNe based on the Galactic ccSN rate and the absence of any
neutrino detections of core collapse (\citealt{Adams2013}).  Stronger priors could
be developed based on the detections of ccSNe progenitors
(e.g., \citealt{Smartt2009}), limits on the rate of 
failed ccSN (\citealt{Kochanek2008}, \citealt{Gerke2014}), 
the neutron star mass function (\citealt{Pejcha2012}), the diffuse neutrino
background produced by ccSNe (\citealt{Lien2010}, \citealt{Lunardini2009}) 
or comparisons of the star and ccSN rates (\citealt{Horiuchi2011},
\citealt{Botticella2012}).   The present results
are consistent with the available constraints, and roughly
predict the mass range and fraction of failed ccSNe needed
to explain the red supergiant problem.

\section*{Acknowledgments}
We thank T. Sukhbold and S. Woosley for sharing the helium core
masses and compactnesses for their expanded set of progenitor 
models.  We thank J.F.~Beacom, D.~Clausen, A.~Gould, C.~Ott, T.~Piro, 
K.Z.~Stanek and T.A.~Thompson for comments and discussions.


\begin{thebibliography}{}
\bibitem[\protect\citeauthoryear{Adams et al.}{2013}]{Adams2013} Adams, S.~M., Kochanek, C.~S., Beacom, J.~F., Vagins, M.~R., \& Stanek, K.~Z.\ 2013, \apj, 778, 164 
\bibitem[\protect\citeauthoryear{Bailyn et al.}{1998}]{Bailyn1998} Bailyn, C.~D., Jain, R.~K., Coppi, P., \& Orosz, J.~A.\ 1998, \apj, 499, 367
\bibitem[\protect\citeauthoryear{Botticella et al.}{2012}]{Botticella2012} Botticella, M.~T., Smartt, S.~J., Kennicutt, R.~C., et al.\ 2012, \aap, 537, A132
\bibitem[\protect\citeauthoryear{Burrows}{1987}]{Burrows1987} Burrows, A.\ 1987, Physics Today, 40, 28 
\bibitem[\protect\citeauthoryear{Clausen et al.}{2014}]{Clausen2014} Clausen, D., Piro, A.~L., \& Ott, C.~D.\ 2014, arXiv:1406.4869
\bibitem[\protect\citeauthoryear{Couch}{2013}]{Couch2013} Couch, S.~M.\ 2013, \apj, 775, 35 
\bibitem[\protect\citeauthoryear{Couch \& O'Connor}{2014}]{Couch2014} Couch, S.~M., \& O'Connor, E.~P.\ 2014, \apj, 785, 123 
\bibitem[\protect\citeauthoryear{Dolence et al.}{2013}]{Dolence2013} Dolence, J.~C., Burrows, A., Murphy, J.~W., \& Nordhaus, J.\ 2013, \apj, 765, 110 
\bibitem[\protect\citeauthoryear{Dolence et al.}{2014}]{Dolence2014} Dolence, J.~C., Burrows, A., \& Zhang, W.\ 2014, arXiv:1403.6115 
\bibitem[\protect\citeauthoryear{Farr et al.}{2011}]{Farr2011} Farr, W.~M., Sravan, N., Cantrell, A., et al.\ 2011, \apj, 741, 103
\bibitem[\protect\citeauthoryear{Fryer et al.}{2012}]{Fryer2012} Fryer, C.~L., Belczynski, K., Wiktorowicz, G., et al.\ 2012, \apj, 749, 91
\bibitem[\protect\citeauthoryear{Gerke et al.}{2014}]{Gerke2014} Gerke, J., et al., 2014, in preparation
\bibitem[\protect\citeauthoryear{Hanke et al.}{2012}]{Hanke2012} Hanke, F., Marek, A., M{\"u}ller, B., \& Janka, H.-T.\ 2012, \apj, 755, 138 
\bibitem[\protect\citeauthoryear{Horiuchi et al.}{2011}]{Horiuchi2011} Horiuchi, S., Beacom, J.~F., Kochanek, C.~S., et al.\ 2011, \apj, 738, 154
\bibitem[\protect\citeauthoryear{Janka et al.}{2008}]{Janka2008} Janka, H.-T., M{\"u}ller, B., Kitaura, F.~S., \& Buras, R.\ 2008, \aap, 485, 199 
\bibitem[\protect\citeauthoryear{Kitaura et al.}{2006}]{Kitaura2006} Kitaura, F.~S., Janka, H.-T., \& Hillebrandt, W.\ 2006, \aap, 450, 345 
\bibitem[\protect\citeauthoryear{Kiziltan et al.}{2013}]{Kiziltan2013} Kiziltan, B., Kottas, A., De Yoreo, M., \& Thorsett, S.~E.\ 2013, \apj, 778, 66 
\bibitem[\protect\citeauthoryear{Kochanek et al.}{2008}]{Kochanek2008} Kochanek, C.~S., Beacom, J.~F., Kistler, M.~D., et al.\ 2008, \apj, 684, 1336
\bibitem[\protect\citeauthoryear{Kochanek}{2014}]{Kochanek2014} Kochanek, C.~S.\ 2014, \apj, 785, 28 
\bibitem[\protect\citeauthoryear{Kreidberg et al.}{2012}]{Kreidberg2012} Kreidberg, L., Bailyn, C.~D., Farr, W.~M., \& Kalogera, V.\ 2012, \apj, 757, 36
\bibitem[\protect\citeauthoryear{Li et al.}{2011}]{Li2011} Li, W., Leaman, J., Chornock, R., et al.\ 2011, \mnras, 412, 1441
\bibitem[\protect\citeauthoryear{Lien et al.}{2010}]{Lien2010} Lien, A., Fields, B.~D., \& Beacom, J.~F.\ 2010, \prd, 81, 083001
\bibitem[\protect\citeauthoryear{Lovegrove \& Woosley}{2013}]{Lovegrove2013} Lovegrove, E., \& Woosley, S.~E.\ 2013, \apj, 769, 109
\bibitem[\protect\citeauthoryear{Lunardini}{2009}]{Lunardini2009} Lunardini, C.\ 2009, Physical Review Letters, 102, 231101
\bibitem[\protect\citeauthoryear{Nadezhin}{1980}]{Nadezhin1980} Nadezhin, D.~K.\ 1980, \apss, 69, 115
\bibitem[\protect\citeauthoryear{Nordhaus et al.}{2010}]{Nordhaus2010} Nordhaus, J., Burrows, A., Almgren, A., \& Bell, J.\ 2010, \apj, 720, 694 
\bibitem[\protect\citeauthoryear{O'Connor \& Ott}{2011}]{Oconnor2011} O'Connor, E., \& Ott, C.~D.\ 2011, \apj, 730, 70
\bibitem[\protect\citeauthoryear{Orosz et al.}{2011}]{Orosz2011} Orosz, J.~A., McClintock, 
J.~E., Aufdenberg, J.~P., et al.\ 2011, \apj, 742, 84 
\bibitem[\protect\citeauthoryear{{\"O}zel et al.}{2010}]{Ozel2010} {\"O}zel, F., Psaltis, D., Narayan, R., \& McClintock, J.~E.\ 2010, \apj, 725, 1918
\bibitem[\protect\citeauthoryear{{\"O}zel et al.}{2012}]{Ozel2012} {\"O}zel, F., Psaltis, D., Narayan, R., \& Santos Villarreal, A.\ 2012, \apj, 757, 55
\bibitem[\protect\citeauthoryear{Pejcha et al.}{2012}]{Pejcha2012} Pejcha, O., Thompson, T.~A., \& Kochanek, C.~S.\ 2012, \mnras, 424, 1570
\bibitem[\protect\citeauthoryear{Smartt et al.}{2009}]{Smartt2009} Smartt, S.~J., Eldridge, J.~J., Crockett, R.~M., \& Maund, J.~R.\ 2009, \mnras, 395, 1409
\bibitem[\protect\citeauthoryear{Sukhbold \& Woosley}{2014}]{Sukhbold2014} Sukhbold, T., \& Woosley, S.~E.\ 2014, \apj, 783, 10 
\bibitem[\protect\citeauthoryear{Takiwaki et al.}{2012}]{Takiwaki2012} Takiwaki, T., Kotake, K., \& Suwa, Y.\ 2012, \apj, 749, 98 
\bibitem[\protect\citeauthoryear{Thompson et al.}{2003}]{Thompson2003} Thompson, T.~A., Burrows, A., \& Pinto, P.~A.\ 2003, \apj, 592, 434 
\bibitem[\protect\citeauthoryear{Ugliano et al.}{2012}]{Ugliano2012} Ugliano, M., Janka, H.-T., Marek, A., \& Arcones, A.\ 2012, \apj, 757, 69
\bibitem[\protect\citeauthoryear{Wong et al.}{2014}]{Wong2014} Wong, T.-W., Fryer, C.~L., Ellinger, C.~I., Rockefeller, G., \& Kalogera, V.\ 2014, arXiv:1401.3032 
\bibitem[\protect\citeauthoryear{Woosley et al.}{2002}]{Woosley2002} Woosley, S.~E., Heger, A., \& Weaver, T.~A.\ 2002, Reviews of Modern Physics, 74, 1015
\bibitem[\protect\citeauthoryear{Woosley \& Heger}{2007}]{Woosley2007} Woosley, S.~E., \& Heger, A.\ 2007, \physrep, 442, 269 
\bibitem[\protect\citeauthoryear{Zhang et al.}{2008}]{Zhang2008} Zhang, W., Woosley, S.~E., \& Heger, A.\ 2008, \apj, 679, 639
\bibitem[\protect\citeauthoryear{Zi{\'o}{\l}kowski}{2014}]{Ziolkowski2014} Zi{\'o}{\l}kowski, J.\ 2014, \mnras, 440, L61 
\end{thebibliography}
\end{document}